\documentclass[aps,prl,twocolumn,superscriptaddress,showpacs]{revtex4-1}

\usepackage{amsmath}
\usepackage{graphicx}
\usepackage{calc}
\usepackage{bm}

\usepackage[T2A]{fontenc}
\usepackage[cp1251]{inputenc}
\usepackage[english]{babel}

\begin{document}

\title{ Surface ferromagnetism in a chiral topological semimetal CoSi.}

\author{N.N. Orlova}
\author{A.A. Avakyants}
\author{A.V. Timonina}
\author{N.N. Kolesnikov}
\author{E.V. Deviatov}

\affiliation{Institute of Solid State Physics of the Russian Academy of Sciences, Chernogolovka, Moscow District, 2 Academician Ossipyan str., 142432 Russia}

\date{\today}

\begin{abstract}
Despite the chiral topological semimetal CoSi is known as bulk diamagnetic, it shows unusual surface ferromagnetism of debatable origin. The ferromagnetic ordering has been attributed to  the distorted bonds, the superlattice of ordered vacancies, or even   to topological surface textures due to the spin polarization in the neighboring Fermi arcs.
We experimentally compare magnetization reversal curves for initially oxidized CoSi single crystals and cleaved samples with a fresh, oxide-free surface. While the oxidized CoSi samples do not show sizable ferromagnetism, the fresh CoSi surface gives a strong ferromagnetic response, which is accompanied by the pronounced modulation of the angle dependence of magnetization, as it can be expected for easy and hard axes in a ferromagnet. In addition to the first order reversal curves analysis, this observation allows us  to distinguish between different mechanisms of the ferromagnetic ordering in CoSi single crystals. We conclude that the surface states-induced RKKY interaction between distorted bonds near the sample surface is responsible for the strong ferromagnetic multi-domain behavior  for freshly cleaved samples.  
\end{abstract}


\maketitle

\section{Introduction}

The concept of a chiral topological semimetal~\cite{bernevig,zhang,chang2017} is usually regarded as modification of Weyl one by additionally broken mirror symmetry. While Weyl semimetals are characterized by multiple pairs of Weyl nodes with small separation in momentum space~\cite{armitage}, there is only one pair of chiral nodes of opposite Chern numbers with large separation in chiral semimetals, which results in extremely long surface Fermi arcs~\cite{long,long1}. Chiral topological semimetals also host new types of massless fermions with a large topological charge~\cite{fermions,chiral_review}, which lead to numerous exotic physical phenomena like unusual magnetotransport~\cite{magnet}, lattice dynamics~\cite{lattice}, and a quantized response to circularly polarized light~\cite{photo}.   

Chiral crystals are known, e.g., as monosilicides of Cr, Mn, Fe, and Co with the simple cubic B20 crystal structure~\cite{MnSi1,MnSi2,Fermi arcs1,review}. Chiral symmetry is provided by 
the neutral 2$_1$ screw axis in the P2$_1$3 space group~\cite{chiral magnetism}. Among these materials, CoSi is the mostly investigated one, so the bulk band structure and the presence of long surface Fermi arcs have been experimentally  confirmed for CoSi  ~\cite{long,cosi1,cosi2,maxChern}. 

One can expect a  complicated response of topological semimetals on the external magnetic field due to the spin-momentum locking and, therefore, spin polarization of topological surface states. For example,  spin polarization of the Fermi arcs in TaAs  lies completely in the plane of the (001) surface and reaches $80\%$~\cite{Xuetal2016}. For the chiral topological semimetal CoSi,    recent theoretical studies have shown that the spin-orbit interaction lifts the spin degeneracy of the surface states leading to their in-plane spin polarization on the (001) surface, with strongly correlated and predominantly antiparallel spin textures in the neighboring Fermi arcs \cite{Burkovetal2018}. In general, spin textures are known in magnetic materials as surface skyrmions~\cite{skyrm1,skyrm2,skyrm3,skyrm4,skyrm5,skyrm6,skyrm7,skyrm8,skyrm9}] or  spin helix structures~\cite{shelix1,shelix2}.

Among the family of chiral crystals, MnSi is a ferromagnet with known low temperature helimagnetic order and skyrmion magnetism near the Curie temperature~\cite{MnSi2}, FeSi is a small-gap semiconductor with an anomalous temperature-dependent magnetic moment~\cite{FeSi}, while CoSi is a diamagnetic semimetal~\cite{CoSidia}. In the latter case, unusual surface ferromagnetism is reported for nanowires,  policrystalline films ~\cite{CoSiNW1,CoSiNW2,CoSialloy}, or even CoSi single crystals~\cite{CoSibulk}. Skyrmion lattice  has been experimentally shown for  the polycrystalline CoSi  films~\cite{review_skyrm,topol_prop,CoSialloy}. Also, unconventional magnon modes have been reported as a joint effect of  surface ferromagnetism and spin-orbit coupling in  CoSi ~\cite{aucosi}.  

For the CoSi surface ferromagnetism, there is no clear understanding of the ordering mechanism.
The Curie point  is near the room temperature (T$_c$ = 328~K), which is the highest one  among all B20-type ferromagnets~\cite{CoSialloy}. One of the mechanisms is due to the distorted bonds: the transition metal (Co) d-orbital electron spin up and spin down populations become asymmetric from the exchange interactions near the CoSi surface~\cite{seo,tai}.
As an alternative, the superlattice of ordered vacancies is considered as a source of ferromagnetic ordering~\cite{CoSiNW2}.  Topological surface textures should also be considered due to the spin polarization in the neighboring Fermi arcs~\cite{Burkovetal2018,chang2022}. Moreover, the surface states-induced RKKY interaction is expected to be available~\cite{Duan45,Zyuzin46,Duan2022} when the topological surface states couple with the lattice-related ordering (e,g, the above mentioned distorted bonds or ordered vacancies). Thus, one should to have a keystone experiment to distinguish between the different proposed mechanisms. 

Here, we experimentally compare magnetization reversal curves for initially oxidized CoSi single crystals and cleaved samples with a fresh, oxide-free surface. While the oxidized CoSi samples do not show sizable ferromagnetism, the fresh CoSi surface gives a strong ferromagnetic response, which is accompanied by the pronounced modulation of the angle dependence of magnetization, as it can be expected for easy and hard axes in a ferromagnet. This observation is compared with the expected one for  different mechanisms of the ferromagnetic ordering in CoSi.

\section{Samples and techniques}

 The initial CoSi material was synthesized from cobalt and silicon powders by $10^\circ $~C/h heating in evacuated silica ampules up to $950^\circ$~С. The ampules were held at this temperature for two weeks and then cooled down to room temperature at 6 $^\circ$~C/h rate. The obtained material was identified as CoSi with some traces of SiO$_2$ by X-ray analysis. Afterward, CoSi single crystals are grown from this initial load by iodine transport in evacuated silica ampules at $1000^\circ$. X-ray diffractometry demonstrates cubic structure of the crystals, also, X-ray spectral analysis confirms equiatomic ratio of Co and Si in the composition, without any SiO$_2$ traces. The quality of our CoSi crystals have been verified in a number of transport experiments~\cite{aucosi,incosi,cosi2w}.
 
To investigate magnetic properties of small CoSi single crystal samples, we use Lake Shore Cryotronics 8604 VSM magnetometer, equipped with nitrogen flow cryostat.  A CoSi crystal is mounted to the sample holder by a low temperature grease, which has been tested to have a negligible magnetic response. 

For the magnetization measurements, we use small (about 2.52-3.91 mg)  CoSi single crystals, see the inset to Fig.~\ref{weak hyster} (a) as an example. Initially, the crystals have been exposed to air (at ambient conditions) for several months, so the surface is covered by the native oxide. After the first step of  magnetization measurements, every crystal has been cleaved to obtain a fresh, oxide-free surface, see the inset to Fig.~\ref{clear hyster} (a). The cleaved sample has been immediately mounted to the flow cryostate for the second step of magnetization measurements.

We obtain hysteresis loops at different temperatures  by the standard method of the magnetic field gradual sweeping between two opposite saturation values. We also perform first order reversal curves (FORC) analysis,  which provides additional information on the magnetic phases and  their interaction. For the FORC analysis the magnetization curves are recorded as a two-dimensional map with the reversal field $H_r$ and  demagnetization field $H$ ~\cite{FORCanalysis,Hr}. Then the obtained $\rho(H,H_r)$ map is usually redrawn in $(H_u,H_c)$ coordinates, where  $H_u = \frac{1}{2}(H + H_r)$ is the interaction field and $H_c = \frac{1}{2}(H - H_r)$ is the coercitivity field, see  Refs.~\cite{CoSnSmag,FORCanalysis} for details.  

The FORC density distribution $\rho(H_u,H_c)$  is known to be convenient for analysis~\cite{FORCtheory,FORCtheory1}. The closed contours of the density distribution peak are usually associated with single-domain regime, while multi-domain material gives open contours that diverge towards the $H_u$ axis. In general, presence of more than  one peak in $\rho(H_u,H_c)$ map corresponds to  multiple magnetic phases. Vertical shift of the peaks  characterizes the type of interaction between the phases:   it is dipolar for the positive shift values while the exchange interaction appears as the negative ones.

\section{Experimental results}

\begin{figure}
\includegraphics[width=1\columnwidth]{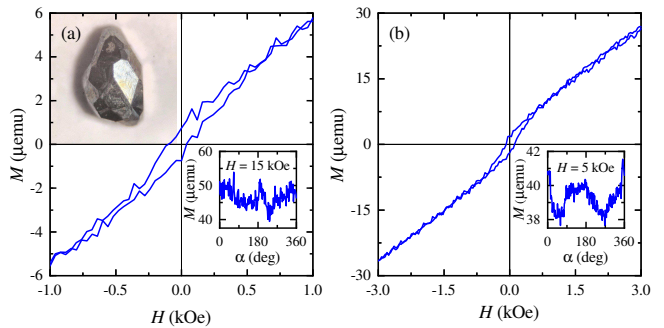}
\caption{(Color online) Magnetization curves $M(H)$  at 100~K temperature for two initially oxidized CoSi crystals, 3.91 mg and 2.52 mg for (a) and (b), respectively. The qualitative behavior is very similar, it well corresponds to the known one~\cite{CoSidia, CoSiNW1,CoSiNW2,CoSialloy, CoSibulk} . The main $M(H)$ dependence corresponds to the  paramagnetic response, $M(H)$ does not demonstrate any clear orientation dependence, as depicted in the insets. The signal amplitude at fixed magnetic field does not scale with the sample mass, so $M(H)$ mostly reflects the contribution of the surface Co oxides~\cite{CoO,Co3O4}, despite a weak hysteresis within $\pm1$~kOe interval.  
}
\label{weak hyster}
\end{figure}

We start magnetization measurements from the initially oxidized samples. Fig.~\ref{weak hyster} shows magnetization loops at 100~K temperature for two   CoSi crystals, 3.91 mg and 2.52 mg for (a) and (b), respectively,  the qualitative behavior is very similar. There is a weak hysteresis within $\pm0.75$~kOe interval of magnetic field.  The main $M(H)$ dependence corresponds to the  paramagnetic response,  which is the known contribution of the surface Co oxides~\cite{CoO,Co3O4}. The sample magnetization $M(H)$ does not demonstrate any specific orientation dependence, as it is shown in the insets to Fig.~\ref{weak hyster}. The signal amplitude at fixed magnetic field does not scale with the sample mass, which also correlates with the surface oxide response.  Thus,  the qualitative $M(H)$  behavior well corresponds to the known one for CoSi semimetal~\cite{CoSidia, CoSiNW1,CoSiNW2,CoSialloy, CoSibulk}.

 As a second step, we repeat the magnetization measurements after cleaving the initial samples from Fig.~\ref{weak hyster} on the smaller fragments with fresh, oxide-free, CoSi surface. 
 
 \begin{figure}[t]
\center{\includegraphics[width=\columnwidth]{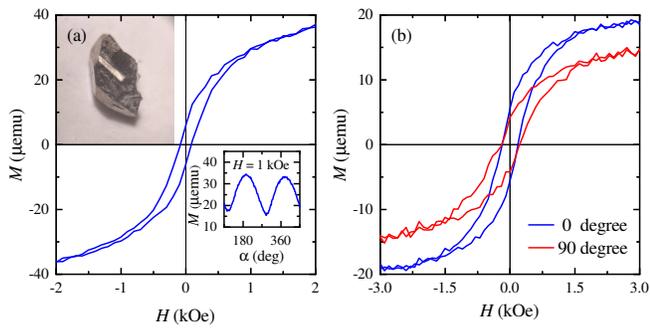}}
\caption{(Color online) Ferromagnetic response from the fresh CoSi surface. The initial samples from Fig.~\ref{weak hyster} are cleaved on  smaller fragments, 1.86 mg and 0.48 mg in  (a) and (b), respectively. Both samples demonstrate one order of magnitude increase in $M$ value, there is now prominent hysteresis  within $\pm2$~kOe field range, which is accompanied by well-defined  $M(H)$ saturation in higher fields. Inset to (a) shows angle dependence of magnetization $M(\alpha)$ with 180$^\circ$ periodic  modulation, reflecting well-defined easy and hard axes in a ferromagnet. For the second sample  in  (b), the angle dependence is shown as two hysteresis loops for these two (easy and hard) field orientations.
}
\label{clear hyster}
\end{figure}
 
To our surprise, both samples demonstrate one order of magnitude increase in $M$ value, while the sample dimensions are diminished to 1.86 mg and 0.48 mg in Fig.~\ref{clear hyster} (a) and (b), respectively. Sample cleaving  has also dramatic  effect on the qualitative $M(H)$ behavior: there is now prominent hysteresis  within $\pm2$~kOe field range, which is accompanied by well-defined  $M(H)$ saturation in higher fields. Thus, the cleaved samples demonstrate standard ferromagnetic response.

Ferromagnetic response from the fresh CoSi surface  is also supported by the angle dependence of magnetization in the inset to Fig.~\ref{clear hyster} (a). We observe strong  modulation of the $M(\alpha)$ with 180$^\circ$ periodicity, as it can be expected for easy and hard axes in a ferromagnet. For the second sample  in Fig.~\ref{clear hyster} (b), the angle dependence is shown as two hysteresis loops for two (easy and hard) field orientations. As usual, the coercivity field is not sensitive to the field direction, while the saturation level varies within 30\% of magnitude, similarly to the angle dependence in the inset to  Fig.~\ref{clear hyster} (a). 

\begin{figure}[t]
\center{\includegraphics[width=\columnwidth]{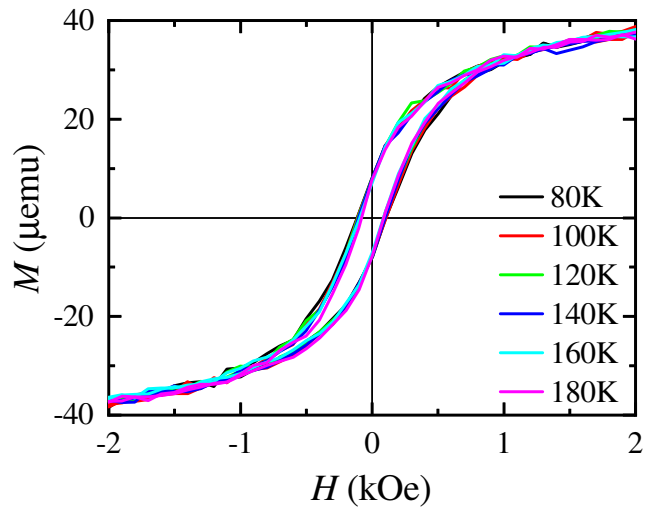}}
\caption{ (Color online) The hysteresis loops for the cleaved CoSi sample from Fig.~\ref{clear hyster} (a), obtained at 80~K, 100~K, 120~K, 140~K,  150~K,  160~K and  180~K temperatures. The curves are practically insensitive to temperature much below the known Curie  point T$_c$ = 328~K for surface ferromagnetism in CoSi~\cite{CoSialloy}. This temperature dependence  confirms the standard ferromagnetic behavior of the fresh CoSi surface.}
\label{M(T)}
\end{figure}

The  Curie  point is above  the room temperature (T$_c$ = 328~K) for surface ferromagnetism in CoSi~\cite{CoSialloy}, so the ferromagnetic response is expected to be practically insensitive to the temperature much below the Curie  point, as we confirm in  Fig.~\ref{M(T)}. The hysteresis loops well coincide from 80~K to 180~K, both the loop width and the saturation level are temperature-independent within this interval. 

\begin{figure}[t]
\center{\includegraphics[width=\columnwidth]{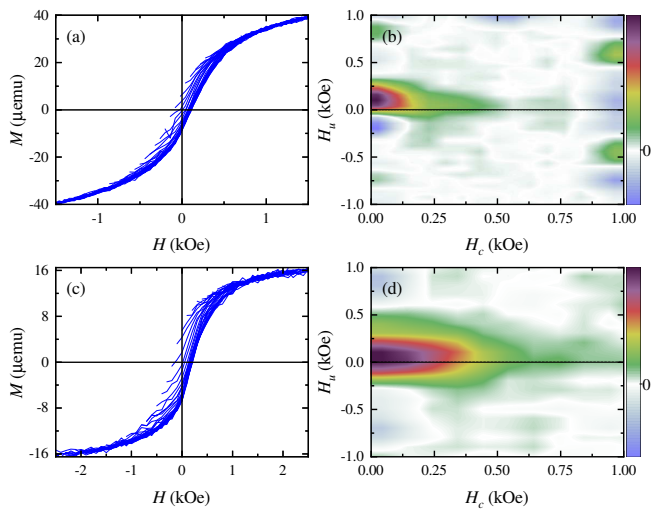}}
\caption{ (Color online) (a,c) The raw FORCs curves for two cleaved samples from  Fig.~\ref{clear hyster} (a) and (b), respectively, at 100~K. (b,d) FORC density diagrams $\rho(H_u,H_c)$ for these samples. Single peak in $\rho(H_u,H_c)$ confirms a single magnetic phase in the sample, while open contours at the $H_u$ axis indicate the multi-domain regime ~\cite{FORCtheory,FORCtheory1}. Thus, a fresh CoSi surface demonstrates strong ferromagnetic multi-domain behavior.
}
\label{FORC}
\end{figure}

Additional information on the ferromagnetic state can be obtained from  FORC data in Fig.~\ref{FORC}. The raw FORC curves and the calculated FORC density diagram $\rho(H_u,H_c)$ are shown in (a,c) and (b,d), respectively, at 100~K temperature for two cleaved samples from Fig.~\ref{clear hyster}. Every  sample shows a single peak in $\rho(H_u,H_c)$, which is centered at low $H_c$ values with so-called open contours at the $H_u$ axis. This behavior is usually regarded as a fingerprint of the multi-domain regime for a ferromagnet~\cite{FORCtheory,FORCtheory1}. The peak center is slightly shifted to the positive values of the interaction field $H_u$, which is  corresponds to the dipolar interaction between domains~\cite{FORCtheory,FORCtheory1}.

\section{Discussion}

As a result, while the samples with oxidized surface show mostly paramagnetic response, the cleaved samples demonstrate strong ferromagnetic multi-domain behavior with definite orientations for easy and hard axes. This behavior is even more surprising, since also the absolute $M(H)$ values are strongly increased for cleaved (i.e. diminished) samples. 

Bulk single crystals of CoSi are known to be diamagnetic with temperature-independent susceptibility~\cite{CoSidia}. This behavior can be confirmed for the initial (oxidized) samples in Fig.~\ref{weak hyster}: after subtracting the paramagnetic $M(H)$ linear dependence in high fields, we obtain   $-0.3*10^6$~emu/g diamagnetic  susceptibility in good correspondence with the previously reported data~\cite{CoSidia}.

On the other hand, the unusual ferromagnetism was observed in CoSi single-crystal nanowires~\cite{CoSiNW1,CoSiNW2}, thin polycrystalline films~\cite{CoSialloy} and  some bulk single crystals~\cite{CoSibulk}. The origin of this ferromagnetism is still debatable. 

(i) Ferromagnetic ordering can  appear due to the distorted and dangling bonds near the sample surface.  The transition metal (Co) d-orbital electron spin up and spin down populations become asymmetric from the exchange interactions near the CoSi surface~\cite{seo,tai}. The unit cell of CoSi contains four Co and four Si atoms, with each Co (Si) atom covalently bonded to six nearest neighboring Si (Co) atoms. In the Co (Si) -- terminated (001) surface, every Co (Si) atom and the underlying Si (Co) one forms a zigzag atomic chain~\cite{Fermi arcs1}. 

(ii) Another source of ferromagnetic ordering is the superlattice of defects (ordered vacancies) in CoSi single crystals~\cite{CoSiNW1,CoSiNW2}. While it is difficult to distinguish experimentally, theoretical simulations suppose that the internal ordered vacancies is the dominant contribution in CoSi single crystal nanowire ferromagnetism.~\cite{CoSiNW2}.  

(iii) Topological surface structures should also be considered for CoSi topological semimetal. 
The spin textures were predicted even for weakly spin-split Fermi surfaces, so the chiral cubic symmetry enforces perfectly parallel spin-momentum locking~\cite{chiral_textures1,chiral_textures2}.  Skyrmion lattice  has been experimentally shown for  polycrystalline CoSi  films~\cite{review_skyrm,topol_prop,CoSialloy}. 

Our experiment indicates, that a fresh surface gives the dominant contribution to the ferromagnetism of  a CoSi single crystal. Thus, the surface effects (i) and (iii) should be considered as the main reason for  strong ferromagnetic  behavior in Fig.~\ref{clear hyster}, while we can not exclude bulk ordered vacancies (ii) as a source of narrow loops in Fig.~\ref{weak hyster}  for the  oxidized samples. To distinguish between the surface ordering mechanisms (i) and (iii), we wish to note that  the structure of the distorted and dangling bonds near the sample surface can be seriously corrupted by the surface oxidation , while the topological effects (iii) are usually considered to have protection from disorder. Thus, the mechanism (i) well correspond to our experiment. On the other hand, one can not expect a strong ferromagnetic response from the distorted  bonds near the sample surface~\cite{seo,tai}.

This apparent inconsistence indicates to a joint effect of the surface-induced ordering mechanisms (i) and (iii).  Naturally, the RKKY interaction is expected to be available for
the surface band of Weyl semimetals, and, therefore, also  for chiral ones~\cite{Duan45,Zyuzin46,Duan2022}. In particular, two distorted bonds can be regarded as magnetic impurities, which are placed on the surface of the semimetal. Considering the spin-exchange interaction (s-d model) between impurities and host electrons, the system Hamiltonian
can be written~\cite{Duan45,Zyuzin46,Duan2022} as 
$$
H=H_{WSM}-J_0 \sum_{i=1,2}\mathbf S_i \mathbf  s_i
$$
where $H_{WSM}$ is the low-energy Hamiltonian of Weyl semimetal,  $J_0$ is the strength of the exchange interaction, $S_i$ is the spin of impurity at site $i$, and $s_i$ refers to the spin of host electrons~\cite{Duan2022}. Mediated by the itinerant host electrons, an indirect exchange interaction (i.e., RKKY interaction) between two impurities is generated, which can be rewritten  in the form of 
$$
H_{RKKY} = J \mathbf S_1 \mathbf S_2
$$
RKKY interaction mediated by surface states in Weyl semimetals  can be induced by different mechanisms~\cite{Duan45,Zyuzin46,Duan2022}. For impurities deposited in the direction perpendicular to the Weyl points splitting, the surface contributions decay much more slowly with impurity distance than that of bulk contribution~\cite{Duan2022}.

Since the topological surface states are confirmed for CoSi ~\cite{long,cosi1,cosi2,maxChern,Burkovetal2018}, the surface states-induced
RKKY interaction can be responsible for the enhancement of the initially weak ferromagnetism of the distorted bonds. The latter is sensitive to the surface oxidation, while the surface states-induced RKKY interaction provides high absolute $M(H)$ values, which are strongly increased for the cleaved (i.e. significantly diminished) samples in Fig.~\ref{clear hyster}. 

As an additional argument, we do not observe the so called bow-tie type hysteresis loops, which are usually ascribed to skyrmions~\cite{Pt/Co/Ta,Pt/Co/Ta2,Co/Pd,Ir/Fe/Co/Pt}.   From the FORC data in Fig.~\ref{FORC}, we should exclude any sizable input from the independent surface phase, since there is only single magnetic phase for the samples with clean crystal surface. This well correlates with the theoretical statement, that the surface states-induced
RKKY interaction survives only when the surface states couple with bulk states (or other surface states of different spins)~\cite{Duan45,Zyuzin46}. This does not contradict to the previous experiments, while  skyrmions have been demonstrated for the polycrystalline samples only~\cite{review_skyrm,topol_prop,CoSialloy}. 

As a result, our experiment allows to distinguish between different mechanisms of the ferromagnetic ordering in CoSi single crystals, so the surface states-induced
RKKY interaction between distorted and dangling bonds near the sample surface is responsible for the strong ferromagnetic multi-domain behavior with definite orientations for easy and hard axes for freshly cleaved samples.

\section{Acknowledgement}

We wish to thank S.S~Khasanov for X-ray sample characterization.

\end{document}